\documentclass[pdflatex,sn-basic]{sn-jnl}

\usepackage{graphicx}
\usepackage{multirow}
\usepackage{amsmath,amssymb,amsfonts}
\usepackage{amsthm}
\usepackage{mathrsfs}
\usepackage[title]{appendix}
\usepackage{textcomp}
\usepackage{manyfoot}
\usepackage{booktabs}
\usepackage{algorithm}
\usepackage{algorithmicx}
\usepackage{algpseudocode}
\usepackage{listings}
\usepackage{float}
\usepackage{tabularx}
\usepackage[section]{placeins}
\usepackage[table]{xcolor}
\usepackage{hyperref}
\hypersetup{bookmarksdepth=3}
\usepackage{tikz}
\usetikzlibrary{positioning, arrows.meta}

\definecolor{highlight}{gray}{0.9}

\definecolor{darkgreen}{rgb}{0.0, 0.5, 0.25}

\theoremstyle{thmstyleone}

\theoremstyle{thmstyletwo}

\theoremstyle{thmstylethree}

\usepackage{etoolbox}
\makeatletter
\newcommand\snj@addressfootnote{%
  \begingroup
    \renewcommand\thefootnote{}%
    \footnotetext{\footnotesize\parindent=0pt\auaddress
      \ifx\corrauthemail\@empty\else *Corresponding author(s). E-mail(s): \corrauthemail\fi}%
  \endgroup
  \global\emailcnt=0 }
\patchcmd{\@maketitle}{\addressfont\auaddress\par}{\snj@addressfootnote}{}{\typeout{Warning: title-block patch not applied}}
\patchcmd{\@maketitle}{\artauthors\par}{{\artauthors\par}}{}{\typeout{Warning: author-line patch not applied}}
\makeatother

\begin{document}

\title[Particle Selection in VR]{Particle Selection in Virtual Reality}

\author*[1,2]{\fnm{Boaz} \sur{Keren-Gil}}\email{krnboa001@myuct.ac.za}
\author[1]{\fnm{James} \sur{Gain}}
\author[3,4,2]{\fnm{Lucia} \sur{Marchetti}}
\author[1]{\fnm{Patrick} \sur{Marais}}
\author[3,2,5]{\fnm{Alexander} \sur{Sivitilli}}

\affil[1]{\orgdiv{Department of Computer Science}, \orgname{University of Cape Town}, \city{Cape Town}, \country{South Africa}}
\affil[2]{\orgdiv{IDIA}, \orgname{Inter-University Institute for Data Intensive Astronomy}, \city{Cape Town}, \country{South Africa}}
\affil[3]{\orgdiv{Department of Astronomy}, \orgname{University of Cape Town}, \city{Cape Town}, \country{South Africa}}
\affil[4]{\orgname{INAF - Istituto di Radioastronomia}, \city{Bologna}, \country{Italy}}
\affil[5]{\orgname{Armagh Observatory and Planetarium}, \city{Armagh}, \state{Northern Ireland}, \country{United Kingdom}}


\abstract{Point-cloud representations of 3D data are used in many fields, from astronomical surveys of galaxies to medical imaging and laser scans of buildings. A task common to all of these is interactive selection: picking out a meaningful subset of points for closer analysis. The data is inherently 3D, yet the traditional workflow uses a 2D mouse on a 2D projection and inherits the limitations of a flat screen. This paper compares Virtual Reality (VR) and desktop environments for 3D particle selection, asking whether VR's six-degree-of-freedom hand interaction yields measurable gains in accuracy, speed, usability, workload and flow. In a counterbalanced within-subjects experiment ($N=34$), participants selected 90 target galaxies from a catalogue of 16,816 galaxies, using mouse-driven selection with keyboard camera control on the desktop and a hand-held spherical brush that adds or removes particles in VR. We measured task accuracy, completion time, and three standard questionnaires: the System Usability Scale (SUS), the NASA Task Load Index (NASA-TLX) and the Flow State Scale (FSS). Accuracy was comparable (91.6\% VR vs.\ 93.1\% desktop; $p=0.858$), but every other measure favoured VR. Tasks were 23\% faster ($p=0.001$, $d=0.93$), usability was higher (SUS 84.6 vs.\ 54.0; $d=-1.78$), workload was lower (NASA-TLX 20.4 vs.\ 34.7; $d=0.90$), and flow was higher (FSS 83.2 vs.\ 70.2; $d=-1.23$). Because the underlying problem, making a 3D selection through a 2D interface, is not specific to astronomy, we expect the pattern to carry over to point-cloud selection in other fields.}

\maketitle

\section{Introduction}\label{sec1}

Scientific and industrial fields increasingly rely on point-cloud representations to capture the 3D structure of real-world data. A particle dataset is a collection of points in space, each with a minimum of 3 spatial coordinates in addition to other associated attributes, such as colour. This is a natural representation when the data does not have continuous surfaces or volumetric density. Such datasets occur across many domains. In medicine, anatomical structures such as the chambers of the heart are reconstructed as point clouds from MRI scans \citep{Beetz2023}. In architecture and cultural heritage, LiDAR surveys produce dense point clouds used for as-built documentation and 3D reconstruction \citep{Vacca2023}. In autonomous systems and urban planning, point clouds underpin applications such as scene understanding and infrastructure management. A task common to all of these is \textit{selection}: picking out a meaningful subset of points for closer analysis, measurement, or processing. 
In each case the analyst's intent is spatial: the region of interest is a volume in the data, not a set of pixels on a screen. Conventional tools nevertheless ask the user to express that intent by moving a mouse across a plane while looking at a projection, so the input, the display, and the structure being operated on are three different spaces that the user must reconcile.

Astronomy provides a particularly demanding instance of this problem, and is the one we study here. Astronomical surveys deal with datasets of significant volume, complexity, and dimension\citep{Zhang2015}. Particles represent galaxies, stars, or other objects, each with coordinates and attributes such as redshift, luminosity, and mass. Beyond observational catalogues, cosmological simulations generate particle datasets in which the particles are the physical constituents themselves---gas, dust, dark matter, and stars---each carrying its own attributes, as in the IllustrisTNG \citep{Nelson2019TNG} and SIMBA \citep{Dave2019SIMBA} simulations. Isolating a subset of these particles for focused analysis is a critical, recurring task, and the lack of genuine depth perception on a 2D interface can limit both accuracy and efficiency.

Several desktop tools exist for particle data, and 2D-to-3D projection limitations appear in all of them. In astronomy, software interfaces with TOPCAT \citep{Taylor2017} and Partiview \citep{Levy2003} provide 2D or limited 3D visualisation. Both are powerful for plotting and tabular analysis, but both require users to interact with projections of 3D data through 2D input devices. Python packages such as Matplotlib \citep{Hunter2007} and Plotly \citep{plotly} add interactivity but share the same shortfall, and CARTA \citep{Wang2026} is designed for images and data cubes rather than 3D point data.  The same pattern shows up outside astronomy. In medical imaging, 3D Slicer \citep{Fedorov2012} is the de facto open-source platform for segmenting volumetric scans and their derived point clouds. In LiDAR and geospatial work, CloudCompare \citep{Girardeau-Montaut2006} provides interactive segmentation, registration, and measurement on dense laser-scanned point clouds. All of these tools share the same underlying constraint: however sophisticated the analysis features are, the user still interacts with 3D data through a 2D mouse on a 2D screen, which is often cumbersome and inaccurate because of the lack of genuine depth perception \citep{Ware2020}.

This paper explores whether Virtual Reality (VR) can overcome these limitations. With stereoscopic depth perception and six degrees of freedom (6-DOF) interaction, VR ought to enhance the selection of particle data. We present the design, implementation, and evaluation of iDaVIE-p, a specialised mode within the iDaVIE software suite \citep{Sivitilli2026, Jarrett2024}, built as a dual-platform tool so that we can compare VR and desktop workflows directly. Although the study uses astronomical data, the selection task (identifying a spatially coherent subset of points in a larger 3D catalogue) is recurrent across the domains mentioned above, so our findings should be transferable. Our contribution is threefold: (1) the design and implementation of a volumetric brush that adds or removes particles with hand-held spheres, (2) a within-subjects comparison of VR and desktop selection across objective and subjective metrics, and (3) evidence that VR significantly improves efficiency, usability, and user experience for inherently 3D particle selection while maintaining comparable accuracy.

\section{Related Work}\label{sec2}
This section reviews selection as an interaction task and the input modalities proposed for it (Section~\ref{subsec21}), the perceptual cues that make a discrete particle's depth legible (Section~\ref{subsec22}), selection techniques for VR and dense point data (Section~\ref{subsec23}), and VR in astronomy, our application domain (Section~\ref{subsec24}).

\subsection{Selection and Interaction}\label{subsec21}
Selection is a fundamental interaction task, often referred to as \textit{manipulation} in the human-computer interaction (HCI) literature. Fitts' Law gives a well-established model of pointing efficiency, linking target acquisition time to distance and size, and it has been extended to three-dimensional targets in consumer VR, where models that account for target depth predict movement time better \citep{Clark2020}. Moving to 3D point cloud data such as astronomical particle catalogues adds complexity from depth and occlusion. Efficient selection in 3D requires navigating along the Z-axis, where some form of depth perception is needed to differentiate targets along a line of sight.

The design of a selection mechanism depends on the user's physical capabilities, the context of use, and the technological constraints. Beyond the standard mouse and controller, recent work has explored gestural interfaces, eye tracking, and brain-computer interfaces (BCIs) as more natural input \citep{Mridha2021}. Eye gaze as a pointer is fast, and combining it with head movement makes it a practical selector in headsets \citep{Sidenmark2019}, while BCIs promise hands-free interaction for accessibility. However, each of these alternatives is a poor fit for the task we study here, which involves selecting potentially thousands of small 3D particles in rapid succession from a dense set. Bare-hand gestures tracked by camera, without a held device, could suffer from imprecision and arm fatigue over the duration scientific analysis requires \citep{Hansberger2017}. Eye gaze introduces the Midas Touch problem \citep{Jacob1990}, where every fixation risks an unintended selection, which is particularly damaging when neighbouring particles are only a tiny fraction of a degree apart in the user's view. BCIs remain low-bandwidth and high-latency relative to controller input, and are not yet practical for bulk interactive selection. Our work therefore builds on the direct-manipulation tradition: a hand-held controller whose position maps one-to-one onto a spatial selection volume in the data, trading novelty for throughput (how much can be selected per minute) and precision. The brush we describe is also operated in mid-air, but it differs from bare-hand gestural input in two ways that matter for precision: activation is gated by a physical trigger rather than inferred from a hand pose, and the tracked controller gives the hand a stable grip and a physical reference. The fatigue cost of holding an arm in the data does remain, and it is one of the design considerations addressed in Section~\ref{subsec31}.

\subsection{Perception of Particle Positions}\label{subsec22}
Depth perception relies on binocular cues (stereopsis, convergence), monocular cues (motion parallax, occlusion, shading), and focus cues (accommodation). Stereopsis exploits the disparity between the images projected on each eye to gauge distance, while convergence refers to the inward movement of the eyes when focusing on nearer objects \citep{Ware2020}. Evidence suggests the brain is not effective at incorporating convergence information unless the target is within arm's length \citep{Viguier2001}, though perceived distance in virtual environments can be recalibrated through experience \citep{CreemRegehr2023}.

Desktop visualisation is limited to monocular cues for simulating depth. Most such cues do not help much when the points are small, discrete objects with no surface detail to exploit \citep{Ware2020}. When the points lie on a real surface, as in LiDAR scans or MRI-derived shapes, a surface can be drawn through them to add shading cues. Individual galaxies or stars have no such surface or orientation, and imposing one risks suggesting structure that is not in the data. Structure-from-motion (kinetic depth) and stereoscopic depth therefore remain the main depth cues for such particle data. A recent perceptual study of 3D point clouds found that rotation revealed hidden structure that stayed invisible in 2D projections and under translation \citep{Gruchalla2021}. Artificial spatial cues can reduce the remaining ambiguity. Proximity luminance covariance, where distant objects appear dimmer, can significantly improve a user's ability to judge relative positions \citep{Wanger1992, Dosher1986}. Shadow cues applied to a whole point cloud can reveal its boundary \citep{Wanger1992}, and artificial spatial cues such as ground-plane lines can help users perceive the depth of individual objects \citep{Ware2020}.

Cues are not weighted equally across tasks \citep{Ware2020}, and some conflict: in conventional VR headsets, the two eyes' images place an object at its simulated distance, but the eyes must still focus on a screen at a fixed distance. This mismatch, the vergence--accommodation conflict, can impair visual performance and increase visual fatigue \citep{Hoffman2008}. Distance is also systematically underestimated in head-mounted displays, although the effect has shrunk as displays have improved \citep{CreemRegehr2023}. For dense particle selection, these limitations mean that depth judgments should not depend on a single cue alone.

\subsection{Selection in VR}\label{subsec23}
A variety of 3D selection techniques have been developed for VR, each trading off reach, precision, and effort \citep{Besancon2021}. The simplest is the Virtual Hand or ``grab'' metaphor, where the controller acts as a one-to-one mapped hand and selection occurs on contact. It is intuitive but bounded by physical reach, so distant data requires a locomotion system. Ray-casting extends reach by emitting a virtual ray from the controller, but jitter at distance degrades accuracy on small targets, such as individual particles. Refinements such as RayCursor \citep{Baloup2019} partially compensate through adaptive snapping and user-controlled ray length. Cone Selection \citep{Steinicke2006} widens the ray into a cone to improve acquisition of small distant targets at the cost of precision on nearby ones. The Go-Go technique \citep{Poupyrev1996} non-linearly extends the virtual arm to combine grab-style directness with longer reach, but overcoming the mismatch between physical and virtual hand positions takes practice. Head-gaze and eye-gaze selection \citep{Blattgerste2018, Piumsomboon2017} offer hands-free alternatives but inherit the Midas Touch problem \citep{Jacob1990}, and eye-gaze selection presupposes eye-tracking hardware that many consumer headsets, including the Meta Quest 2 used here, do not have. We avoid depending on it so that the tool remains deployable on the commodity hardware most research groups own.

More recent work has tackled selection in dense 3D environments such as point clouds. \citet{MontanoMurillo2020} introduced Slicing-Volume, a hybrid technique where a mid-air 3D volume is projected onto a physical tablet for fine-grained 2D refinement, and showed that adding a touchable surface improves accuracy in highly occluded scenes. \citet{Liebers2023} compared desktop, tablet, and VR for manual segmentation of 3D point clouds, and found that while VR offered higher subjective engagement and immersion, desktop and tablet interfaces outperformed VR on task completion time, highlighting an ongoing trade-off between immersive spatial understanding and interaction efficiency. \citet{Doula2022} built Immersive-Labeler, a VR environment for annotating large urban point clouds; rather than proposing a new selection primitive, it pairs immersive navigation with object-based visual aids, and a study with 20 participants found that higher immersion together with these aids made annotation faster and more engaging. The most recent systems focus on scale. \citet{Lin2024} label clouds of up to 50 million points at VR frame rates with consumer controllers, using a mixed CPU/GPU data structure for immediate feedback; \citet{PereiraCosta2026} push this to over 150 million points with a labelling volume that adapts to the local shape of the cloud; and \citet{Monica2026} label a few key points with a hand or gaze ray and let the labels spread to neighbouring points along the shortest paths between them, an approach that beat a bounding-box baseline on both time and quality. A common thread in these works is that bounding-box or bounding-volume selection remains the dominant VR metaphor for point clouds \citep{Wirth2019, Liebers2023, Monica2026}, which typically requires the user to position, rotate, and scale a discrete widget for each target region.

Our work departs from the bounding-volume paradigm in favour of a continuous, brush-style volumetric selector with boolean addition and subtraction operations. Rather than placing and sizing a box, the user sweeps a hand-held sphere through the data, selecting incrementally. This is better matched to workflows in which target regions are irregular in shape and do not fit neatly inside a box, and where the user has to refine selections repeatedly across diverse clusters. The volumetric brush keeps the fidelity of direct manipulation, avoids the precision issues of ray-casting on small targets, and does not depend on the eye-tracking hardware that most headsets lack.

\subsection{VR in Astronomy}\label{subsec24}
VR in astronomy has expanded how researchers visualise and interact with data \citep{Ferrand2016, Fluke2018, Lan2021}. VR has been used to measure the 3D structure of supernova-remnant ejecta, exploiting human depth perception to characterise complex structures quickly \citep{Baracaglia2020}, and 3D astronomical models are increasingly delivered through printing and extended reality \citep{Arcand2020}; Gaia Sky renders the positions and motions of over a billion stars from the Gaia catalogue with VR output \citep{Sagrista2019}. FRELLED is an open-source Blender toolkit for inspecting 3D radio data cubes in real time, which can now export them for viewing in VR headsets \citep{Taylor2025}. More broadly, VR and Augmented Reality are being explored for mission planning, operations, and high-dimensional data visualisation in astrophysics \citep{Jarrett2021}. Multi-wavelength images and volumetric cubes, combined in VR, allow astronomers to explore cosmic systems more intuitively \citep{Marchetti2020}. Together these show that VR in astronomy serves as more than a novelty; it enables new forms of data interrogation and scientific discovery.

The present work builds directly on the iDaVIE software suite \citep{Jarrett2024}, which is open source~\citep{Sivitilli2026} and publicly available~\footnote{\url{https://github.com/idia-astro/iDaVIE}}. 
The suite comprises two primary modes: iDaVIE-v, for volumetric rendering of 3D data cubes, on which the public release is based, and iDaVIE-p, for particle and catalogue rendering, which is included in the open-source repository but not yet in the packaged public release. Before this work, iDaVIE-p supported visualisation but not selection. The present study contributes the selection functionality described in Section~\ref{sec3}, together with an octree spatial index that keeps its queries fast for large particle counts.

\section{Design}\label{sec3}
This section sets out the design goals agreed with the iDaVIE team (Section~\ref{subsec31}), the VR implementation iDaVIE-p, including the selection metaphors considered, the depth cues used, and the data structures behind the brush (Section~\ref{subsec32}), and the desktop baseline, which shares the rendering pipeline so that the comparison isolates the interaction paradigm (Section~\ref{subsec33}).

\subsection{Design Goals}\label{subsec31}
The design of the selection mechanism followed HCI goals tailored to scientific data analysis, defined with the iDaVIE development team. Foremost is accuracy: false positives (unwanted background particles) can be as damaging as false negatives. Efficiency is also critical, since users need to isolate large clusters without lag. Given the physical nature of VR, the method must minimise fatigue, including gorilla arm syndrome (the tiredness of holding the arms raised for long periods) \citep{Hansberger2017}, and cognitive load, and should feel natural, ideally mimicking real-world actions like painting. Because scientific analysis is iterative, the system must support rapid selection and deselection adjustments without a full restart, following the principles of direct manipulation \citep{Shneiderman1983}. It further requires scalability from hundreds to millions of particles, error correction in line with Nielsen's heuristic of user control and freedom \citep{Nielsen1994}, and a low barrier to entry so that researchers without VR experience can learn it quickly.

\subsection{VR Particle Selection: iDaVIE-p}\label{subsec32}
We developed iDaVIE-p selection in the Unity engine, building on the iDaVIE platform \citep{Sivitilli2026}. The software renders astronomical source catalogues (galaxies, stars, and so on) as particles in a 3D environment, but it is context agnostic and does not impose a meaning on the particles.

Selection in a boundless 3D environment requires robust locomotion. iDaVIE uses a two-handed grip metaphor in the tradition of direct manipulation of 3D data \citep{Cutler1997}. Pressing the grip trigger translates the dataset. Moving the controllers apart or together while gripping zooms in or out. Pressing both grip triggers and moving one controller around the other rotates the dataset. This lets users move quickly from a god's-eye view of the catalogue to a localised view for precise selection.

We considered two selection metaphors: a spatial deformation cylinder and a sphere-based volumetric brush. Both were prototyped on a mock dataset in Blender with three difficulty levels: easy (particles on the outer edge), medium (particles in the centre, requiring rotation), and hard (particles obscured behind non-targets).

The \textit{Spatial Deformation Cylinder} encloses particles in a generalised cylinder (a tube whose path and width can vary along its length), whose shape can be modified using deformation handles \citep{Gain2008}. It offers strong scalability (one object selects many particles) and few iterations, but ranked poorly for speed and undo/adjust: fitting a cylinder to an amorphous cluster requires meticulous manipulation of many control points, which is cognitively demanding, physically tiring, and presupposes 3D-design expertise most users lack.

The \textit{Sphere-Based Brush} is our chosen metaphor. It uses spheres for selection and deselection, mimicking a 3D paint tool. We picked it after a systematic comparison across all the design goals, in consultation with the iDaVIE team. Each goal was ranked from 1 (worst) to 5 (best) for each method, so a higher score always means the method does better on that goal: a high score on fatigue means the method is less tiring, and a high score on required expertise means less expertise is needed. The sphere-based brush scored higher on fatigue (4 vs.\ 2), fidelity (4 vs.\ 2), speed (5 vs.\ 2), required expertise (5 vs.\ 2), and undo/adjust (5 vs.\ 1). The cylinder scored higher only on iteration (5 vs.\ 1). Both scored equally on accuracy, dimensionality, scalability, and implementation complexity.

Our chosen sphere-based tool design has two modes:
\begin{itemize}
    \item \textit{Select mode}: The user holds the controller's trigger to spawn selection spheres. Any particle within the sphere volume is added to the selection set. This acts as a 3D paint brush, letting users sweep through clusters of data.
    \item \textit{Deselect mode}: Pressing the A button on the right controller switches the brush to deselection, and pressing it again switches back. Enclosing spheres in this mode remove particles from the selection, acting as a 3D eraser. This supports the undo/adjust goal.
\end{itemize}
Each sphere action is atomic, so a mistaken sweep is corrected locally with the opposite mode rather than by restarting the selection.

\textbf{Depth cues for selection.} iDaVIE-p renders against a black background, the convention in astronomical visualisation: it mirrors the night sky, keeps faint particles visible, and keeps the distance colour mapping legible. The tradeoff is that it removes depth cues that come from the surroundings, such as distant objects fading into haze and ground-plane lines, so the design relies on cues from the particles themselves, stereoscopic depth and structure-from-motion.

For the iDaVIE-p implementation we prioritised the cues highlighted in Table~\ref{tab:depth_cues}. In VR, depth from stereo vision and from motion comes almost for free, because ordinary head movement makes near particles shift more than far ones. We found that two simple cues, nearer particles hiding farther ones (interposition) and farther particles looking smaller (relative size), were also needed to tell which particle is in front in dense clusters that would otherwise read as a flat wall of points, and we added one artificial cue, a colour map driven by each galaxy's distance from Earth, to substitute for the lack of surface texture. Specifically:
\begin{itemize}
    \item \textit{Colour by distance} (artificial cue): the catalogue colour map is driven by each galaxy's distance from Earth, so hue orders the catalogue in depth. Unlike proximity luminance covariance \citep{Dosher1986}, this ordering does not change with the viewer's position.
    \item \textit{Kinetic depth}: we rejected per-particle animations after testing (gentle X-axis oscillations became confusing when zoomed out) and rely on structure-from-motion derived from head and controller tracking. As the user moves their head or rotates the dataset, the parallax between near and far particles reveals 3D structure.
    \item \textit{Relative size}: standard perspective scaling. More distant particles appear smaller. Future iterations may explore dynamic density rendering, in which the display adapts to how densely the particles are packed.
\end{itemize}
Other cues, such as ground-plane lines \citep{Ware2020}, were considered but rejected to avoid visual clutter in datasets with thousands of active particles.

\begin{table}[!htbp]
\centering
\small
\begin{tabularx}{\textwidth}{l l X}
\toprule
\textbf{Depth cue} & \textbf{Type} & \textbf{Description} \\
\midrule
Binocular disparity & Binocular & The difference in the images received by each eye due to their horizontal separation, giving depth through stereopsis. \\
\addlinespace
Convergence & Binocular & The inward turning of the eyes when focusing on a close object. Depth is cued by the degree of convergence. \\
\addlinespace
Motion parallax & Monocular & Objects closer to the observer move faster across the field of vision than objects further away when the observer moves. \\
\addlinespace
Linear perspective & Monocular & Parallel lines appear to converge in the distance at a vanishing point. \\
\addlinespace
Texture gradient & Monocular & Texture density increases with distance; closer objects show more detail than distant ones. \\
\addlinespace
\rowcolor{highlight} Interposition & Monocular & When one object overlaps another, the overlapping object is perceived as closer. \\
\addlinespace
\rowcolor{highlight} Relative size & Monocular & If two objects are known to be similar in size, the smaller-looking one is perceived to be further away. \\
\addlinespace
Height in field & Monocular & Objects closer to the horizon in the visual field are perceived as further away. \\
\addlinespace
Light and shadow & Monocular & Shading and shadow patterns give cues about distance and contour. \\
\addlinespace
\rowcolor{highlight} Kinetic depth & Monocular & Movement of an object reveals its 3D shape, e.g.\ when a rotating object exposes its structure. \\
\addlinespace
\rowcolor{highlight} Artificial cues & Both & Cues added by the designer rather than present in the scene, such as distance-based dimming (proximity luminance covariance), ground-plane grids, drop lines, or halos, used to make depth legible in rendered 3D data. \\
\bottomrule
\end{tabularx}
\caption{Depth cues in visual perception. Highlighted rows indicate the cues prioritised in the design of iDaVIE-p.}
\label{tab:depth_cues}
\end{table}

\FloatBarrier

To ensure scalability and maintain high framerates during intensive interaction, the particles and each particle's selection state are drawn on the graphics card (GPU) by custom shader programs. Drawing millions of particles on the CPU would be far too slow, so all particle data (ID, position, colour, and shape) is passed to the GPU in a single buffer.

\textbf{Data structure.} Each particle carries a selection state flag with three values, stored alongside the particle's other data in that GPU buffer: \textit{Initial}, \textit{Selected}, and \textit{Unselected}. Before any selection every particle is \textit{Initial} and is drawn at full opacity in its base colour, so that every particle is a clearly visible potential target. The first Select sweep moves the swept particles to \textit{Selected} and, at the same time, moves every other \textit{Initial} particle to \textit{Unselected}. \textit{Selected} particles are drawn with a different sprite (a filled circle in place of the default halo) at full opacity, so the current selection stays visually emphasised; \textit{Unselected} particles are drawn at 20\% of their base opacity, so the user can focus on the selection without losing the surrounding context. A Deselect sweep moves \textit{Selected} particles back to \textit{Unselected}, so a deselected particle is rendered like any other particle that is not part of the current selection. The third state exists because the tool must distinguish ``no selection has been made yet'' from ``not part of the current selection'': dimming everything in the former case would hide the data, whereas in the latter case the dimming is what makes the selection legible. A mistaken deselection is reversed locally by sweeping the Select brush over the affected particles again; once no particle remains \textit{Selected}, every particle automatically returns to \textit{Initial}.

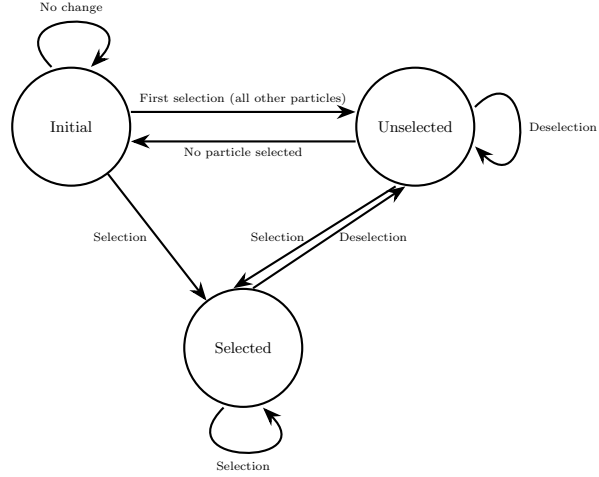
\begin{figure}[!htbp]
\centering
\begin{tikzpicture}[scale=0.65, transform shape,
    >={Stealth[length=2.5mm]},
    state/.style={circle, draw, thick, minimum size=24mm, align=center, font=\small, inner sep=0pt},
    every node/.style={font=\footnotesize}
]

\node[state] (U) at (0, 0)      {Initial};
\node[state] (D) at (7, 0)      {Unselected};
\node[state] (S) at (3.5, -4.5) {Selected};

\draw[->, thick] (U) -- node[midway, left=2pt] {Selection} (S);

\draw[->, thick] ([xshift=2mm]S.north) -- node[midway, right=2pt] {Deselection} ([xshift=-2mm]D.south);

\draw[->, thick] ([xshift=-4mm]D.south) -- node[midway, left=2pt] {Selection} ([xshift=-2mm]S.north);

\draw[->, thick] ([yshift=3mm]U.east) -- node[midway, above] {First selection (all other particles)} ([yshift=3mm]D.west);

\draw[->, thick] ([yshift=-3mm]D.west) -- node[midway, below] {No particle selected} ([yshift=-3mm]U.east);

\draw[->, thick] ([xshift=-4mm]U.north) .. controls +(-1.2,1.2) and +(1.2,1.2) .. ([xshift=4mm]U.north)
    node[midway, above, yshift=2pt] {No change};

\draw[->, thick] ([yshift=4mm]D.east) .. controls +(1.2,1.2) and +(1.2,-1.2) .. ([yshift=-4mm]D.east)
    node[midway, right, xshift=2pt] {Deselection};

\draw[->, thick] ([xshift=-4mm]S.south) .. controls +(-1.2,-1.2) and +(1.2,-1.2) .. ([xshift=4mm]S.south)
    node[midway, below, yshift=-2pt] {Selection};

\end{tikzpicture}
\caption{State diagram for the three-state selection flag. Initial particles are drawn at full opacity in their base colour, Selected particles with a filled-circle sprite in place of the default halo, and Unselected particles at 20\% opacity. The first Select sweep also moves every other Initial particle to Unselected; once no particle remains Selected, all particles return to Initial.}
\label{fig:selection_states}
\end{figure}
\textbf{Selection logic.} Interaction is driven by a volumetric brush in the shape of a resizeable sphere. When the user activates the brush, a selection sphere of radius $S_r$ is placed at the brush position, and every particle whose centre lies within $S_r$ of the sphere's centre is selected. Comparing squared distances avoids slow square-root calculations. Particles are indexed in an octree built once when the catalogue loads, so the test runs only on particles in nodes that intersect the sphere; the octree changes how quickly a query is answered, not which particles it selects. The selection radius is exactly the radius of the rendered preview sphere: a particle is flagged as soon as its centre lies inside the visible brush, so the volume the user sees is the volume that selects. Each selection sphere stays visible for about a second and then fades out, so rapid brushing does not clutter the view.

\begin{figure}[!htbp]
    \centering
    \includegraphics[width=0.6\linewidth]{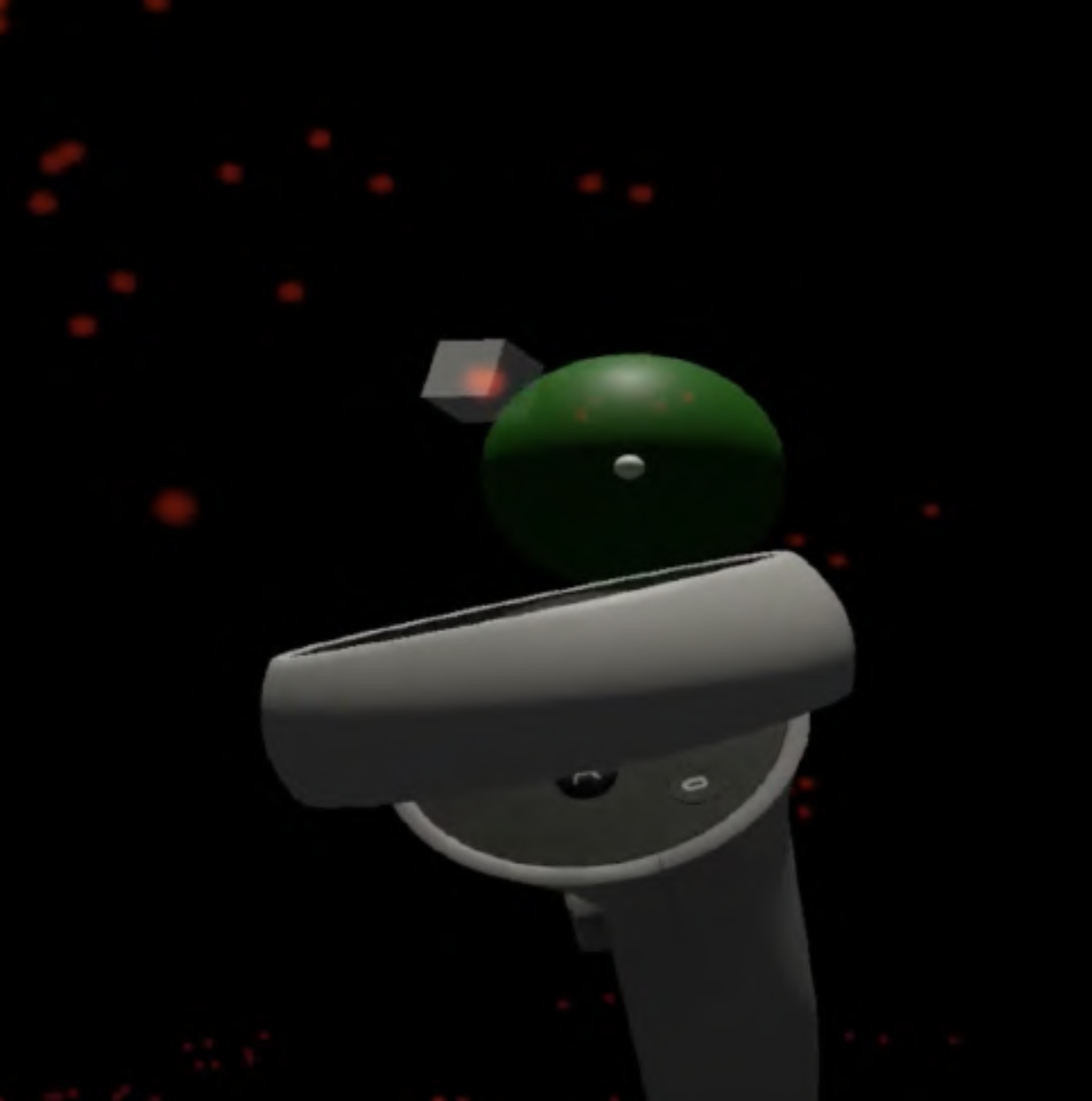}
    \caption{The selection brush in iDaVIE-p at close range. The preview sphere attached to the controller shows where the next selection sphere will be spawned; the white bounding boxes mark the target particles used in the experiment.}
    \label{fig:idaviep_interface_2}
\end{figure}

The sphere radius is adjustable via the controller joystick, with a minimum threshold to prevent accidental collapse to zero, letting users balance between fine and coarse selection. A preview sphere attached to the controller gives continuous visual feedback of where the next selection sphere will be spawned, with colour indicating the active mode: green for Select, red for Deselect (Figure~\ref{fig:idaviep_interface_2}).

\textbf{Visual feedback.} Feedback comes from the state transitions in the selection flag. In \textit{Select} mode, particles inside the brush move to \textit{Selected} and, on the first sweep, all remaining \textit{Initial} particles are dimmed to \textit{Unselected}. In \textit{Deselect} mode, \textit{Selected} particles inside the brush return to \textit{Unselected}, and particles that were not selected are left untouched. Because the state flag lives in the GPU buffer, the change in rendering (sprite change for \textit{Selected}, 20\% opacity for \textit{Unselected}) appears in the same frame as the brush stroke, so users perceive the spatial extent of their selection in real time. In the experiment, target particles were additionally marked with a white bounding box that turned green when selected (Figure~\ref{fig:idaviep_interface_2}), and were exempt from the dimming so that they remained findable.

\textbf{Rendering pipeline.} Each particle is drawn as a small square that always faces the viewer, textured with its sprite and softened at the edges to avoid pixelation; its size, colour, opacity, and shape come from a settings file or the GPU buffer.

\subsection{Baseline Desktop Application}\label{subsec33}
To provide a valid baseline, we developed a desktop version of iDaVIE-p. The rendering pipeline matches the VR version for visual parity, but the interaction paradigms were adapted to standard keyboard and mouse input to reflect traditional scientific workflows.

Desktop locomotion follows the flying camera controls typical of first-person shooter games and 3D modelling software (for example, Blender). Navigation is split between camera movement and orientation:
\begin{itemize}
    \item \textit{Translation}: arrow keys or the scroll wheel move the camera forward, backward, and sideways.
    \item \textit{Panning}: W, A, S, and D keys pan vertically and horizontally.
    \item \textit{Orientation}: SHIFT plus mouse drag rotates the viewpoint around the dataset; SPACE plus mouse drag pans the view. This redundancy follows common 3D-software conventions to minimise the learning curve.
\end{itemize}

Desktop selection projects 2D screen coordinates into the 3D data space. We implemented two modes:
\begin{enumerate}
    \item \textbf{Single-particle selection.} A click selects or deselects a particle directly via a ray cast from the camera. This is accurate for isolated targets, but does not account for occlusion; users must navigate around dense clusters to reach obstructed particles. Unlike VR selection, the mouse pointer supplies only X and Y screen coordinates; the ray selects the first particle it hits, so the user never specifies depth explicitly. Depth still matters in two ways: a distant particle looks smaller on screen and so demands finer pointing, and when several particles overlap along the line of sight only the nearest can be picked, so the user must move the camera to reach the others.
    \item \textbf{Volumetric selection box.} To approximate the bulk selection capability of the VR spheres, we implemented a 3D selection box. CTRL+Click brings up a semi-transparent cube in the scene. Users manipulate this box using translation and scaling handles on the X, Y, and Z axes, coloured by the standard axis convention (red for X, green for Y, blue for Z): spheres translate the box along their axis and cubes scale it. These handles are interface widgets, not data particles. Unlike the VR brush, which creates an immediate selection volume at the hand's position, the desktop box needs a multi-step process: positioning, adjusting its dimensions with the handles, checking the depth coverage by rotating the camera, and finally pressing ENTER to confirm (or BACKSPACE to cancel). This highlights the inherent difficulty of specifying 3D volumes on a 2D screen.
\end{enumerate}

\begin{figure}[!htbp]
    \centering
    \includegraphics[width=0.6\linewidth]{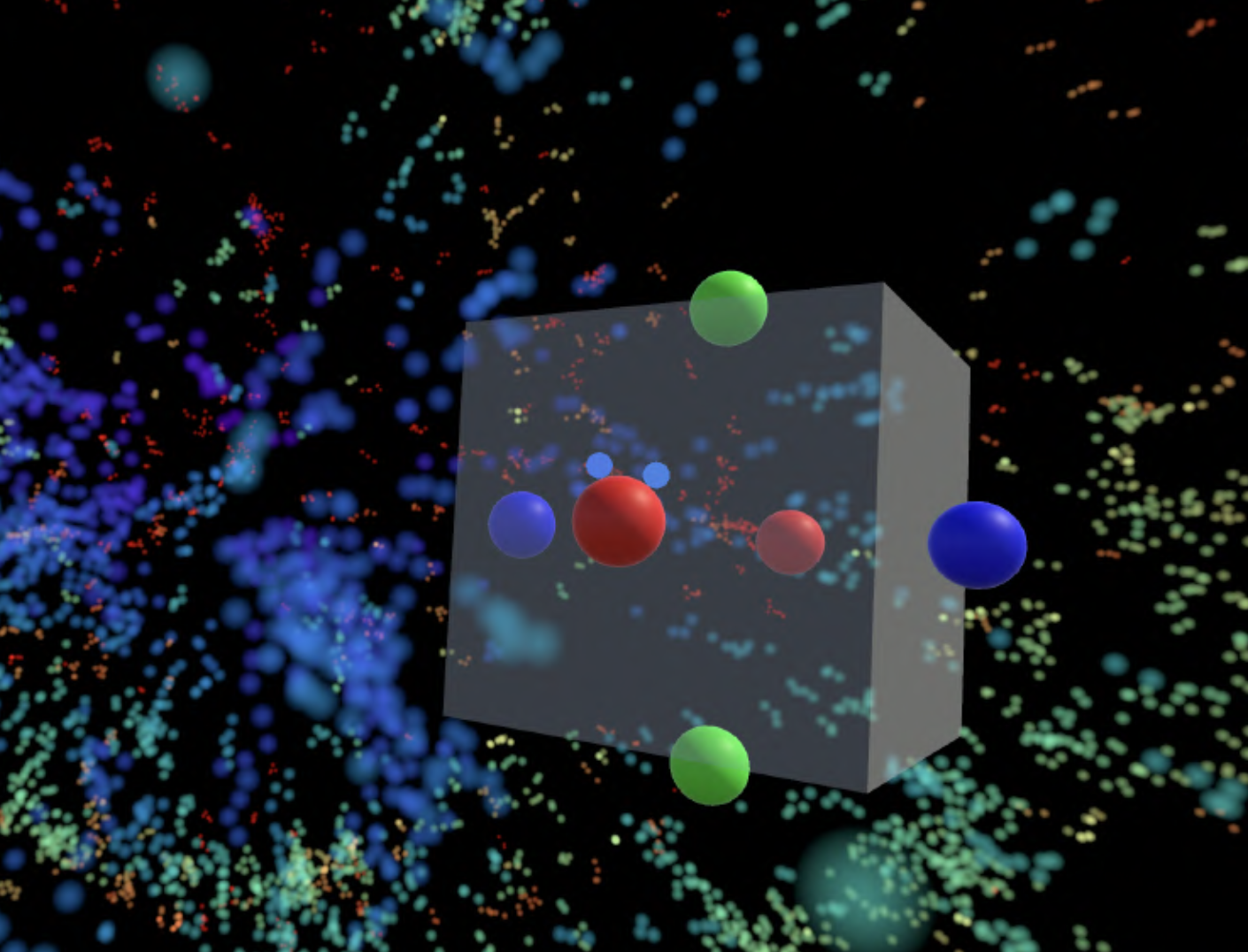}
    \caption{Volumetric selection box applied in the desktop application to select particles. The coloured handles follow the axis convention (red for X, green for Y, blue for Z): spheres translate the box, cubes scale it. They are interface widgets, not data particles.}
    \label{fig:desktop-box}
\end{figure}

\section{Experiment Design}\label{sec4}

This section describes the experiment used to compare the VR and desktop implementations of iDaVIE-p: a within-subjects design in which each of $n=34$ participants performed a particle-selection task on both platforms, using two target sets (Set A and Set B) drawn from the Two Micron All-Sky Survey (2MASS) Redshift Survey~\citep{Huchra2012, Skrutskie2006}, with two counterbalanced groups that differed in platform order and in set-to-platform assignment. Task performance was measured through accuracy and completion time, and user experience through the System Usability Scale (SUS), the NASA Task Load Index (NASA-TLX), and the Flow State Scale (FSS). We set out the hypotheses (\S\ref{subsec41}), the methodology and procedure (\S\ref{subsec42}), the measures (\S\ref{subsec43}), and the statistical analysis (\S\ref{subsec44}).

\subsection{Hypotheses}\label{subsec41}
We defined the following null ($H_0$) and alternative ($H_1$) hypotheses for the platform comparison (VR vs.\ Desktop):
\begin{itemize}
    \item $H_0$: There is no significant difference between desktop and VR in (a) SUS score, (b) NASA-TLX score, (c) FSS score, (d) task completion time, or (e) accuracy of task completion.
    \item $H_1$: There is a significant difference between desktop and VR in the corresponding measure (a)--(e).
\end{itemize}
Each lettered measure is tested separately, so $H_0$ may be rejected for some measures and retained for others.

\subsection{Methodology}\label{subsec42}
Every participant performed the selection task on both platforms, giving a direct paired comparison of task performance and user experience.

\subsubsection{Treatment and Dataset Generation}\label{subsubsec421}
The experiment used the 2MASS Redshift Survey (2MRS), one of the most complete surveys of the local universe, covering 91\% of the sky with roughly 45,000 galaxies \citep{Huchra2012, Jarrett2004}. The catalogue loaded in the study was the 2MRS Galaxy Group Catalogue \citep{Lambert2020}, i.e.\ the 16,816 2MRS galaxies that belong to identified groups, which gives the clustered structure a volumetric selection tool is meant for.

Participants always saw the full 16,816-galaxy catalogue. The task was to find and select a designated set of 90 target particles embedded in it, each marked with a white bounding box, while the remaining 16,726 particles acted as distractors. Two such target sets (Set A and Set B) were generated from the catalogue. The two sets provide, by construction, consistent difficulty across trials and prevent rote learning. Each contained exactly 90 target particles. Of these, 80 were clustered particles generated using a $k$-means algorithm ($k=5$) based on distance from the dataset centroid. We chose $k=5$ with the common ``elbow'' rule: the point beyond which adding further clusters barely makes the groups any tighter. Below $k=5$ the clusters became large and spatially overlapping; above it they fragmented into groups too small to be visually distinct. At $k=5$ the clusters occupied five well-separated regions of the catalogue, which is the kind of shape a volumetric selection tool is designed for: a brush or box should be able to enclose many target particles in a single sweep. So this distribution directly stresses the central design claim of the VR tool. The spatial separation between clusters also forces the user to navigate between selection events, which is the dominant time cost in the task and the component of efficiency we most wanted to measure. The remaining 10 targets were scattered outliers, drawn at random from the 1,000 particles furthest from the centroid. These required users to locate and select individual targets without accidentally capturing background noise, testing fine-grained rather than bulk volumetric selection.

We initially considered 180 targets, but internal testing showed completion within the time limit was nearly impossible, so we reduced the set to 90. Picking cluster particles at random around a centre (Gaussian sampling) was also tried, but it gave sets that looked too alike; $k$-means gave distinct sets, so participants were less likely to recognise them in the second round.

\subsubsection{Participants}\label{subsubsec422}
We recruited 34 participants from the University of Cape Town, mainly students and researchers from the Astronomy (22) and Computer Science (5) departments, with others from various departments. Most had little to no prior VR experience ($n=26$; the remaining 8 reported occasional or frequent use). No participants reported epilepsy or colour blindness. Participants were randomly divided into two equal groups of 17. The grouping was not based on demographics or experience; it served a counterbalancing purpose along two axes. Each group used a different target set on each platform: Group 1 used Set A in VR and Set B on the desktop, while Group 2 used Set A on the desktop and Set B in VR. Every participant therefore saw each set exactly once, and across the two groups each set was used on each platform by 17 participants, so any difficulty specific to one particle layout was absorbed rather than attributed to a platform. The two groups also received the platforms in opposite order (Group 1 VR first, Group 2 Desktop first), which absorbs any learning or fatigue effects that would otherwise accrue to whichever platform was always tested second. Full details of the sequence are given in Section~\ref{subsubsec423}.

The three questionnaires cover complementary dimensions: the SUS captures whether users can operate the interface, NASA-TLX how hard it is to operate, and neither speaks to the sustained, absorbed concentration a scientist needs across a long analysis session. For that we included the FSS \citep{Jackson1996}, which measures flow~\citep{csikszentmihalyi1992}; immersion and proprioceptive mapping (the brush acting exactly where users feel their hand to be) are precisely what VR is supposed to add over a flat screen, and a usability-and-workload assessment alone would miss that effect. An FSS dimension, Autotelic Experience, in fact produced the largest effect size of any subjective measure (Section~\ref{sec5}).

\subsubsection{Procedure}\label{subsubsec423}
The experiment followed a strict protocol:
\begin{enumerate}
    \item \textbf{Pre-experiment.} Participants completed a demographic form and were briefed on the study goals. The form collected date of birth; academic level (undergraduate, honours, MSc, PhD, postdoctoral or research staff, academic staff); department; prior VR experience on a three-point scale (little to none: never, or up to twice a year; some: up to twelve times a year; experienced: more than twelve times a year); frequency of playing computer games (rarely or never, monthly, weekly); and a yes/no screening for epilepsy and colour blindness. For the analysis in Section~\ref{subsec51}, the first VR-experience category is treated as ``novice'' and the other two as ``experienced''.
    \item \textbf{VR playground (15 mins).} Regardless of starting platform, all participants were provided time in a VR Playground with a mock dataset. This let them acclimatise to the head-mounted display (HMD; Meta Quest 2, connected via Oculus Link to a desktop PC meeting SteamVR requirements with an NVIDIA GTX 1070 or equivalent), the locomotion, and the selection controls, mitigating the learning curve during the measured test. Participants were encouraged to ask questions during this phase.
    \item \textbf{Tutorial videos.} Before each platform test, participants watched a tutorial video showing successful task completion on the appropriate platform. The video was recorded on the mock playground dataset rather than on Set A or Set B, so it did not reveal the location of any test target.
    \item \textbf{Counterbalanced tasks (20 mins maximum).} To absorb dataset-layout effects and platform-order effects, the two groups were assigned mirrored sequences:
        \begin{itemize}
            \item Group 1: VR first (Dataset A), then Desktop (Dataset B).
            \item Group 2: Desktop first (Dataset A), then VR (Dataset B).
        \end{itemize}
    This cross-assignment means each dataset was used on each platform by an equal number of participants, and each platform was tested first by an equal number of participants.
    
    Participants selected the 90 target particles (highlighted with a white bounding box that turned green upon selection) as accurately and quickly as possible. Auditory signals accompanied successful and unsuccessful selections. The session was capped at 10 minutes.
    \item \textbf{Post-experiment.} Immediately after each task, participants filled in the feedback questionnaires, thus submitting two sets of all questionnaires.
\end{enumerate}

\begin{figure}[!htbp]
    \centering
    \includegraphics[width=0.7\linewidth]{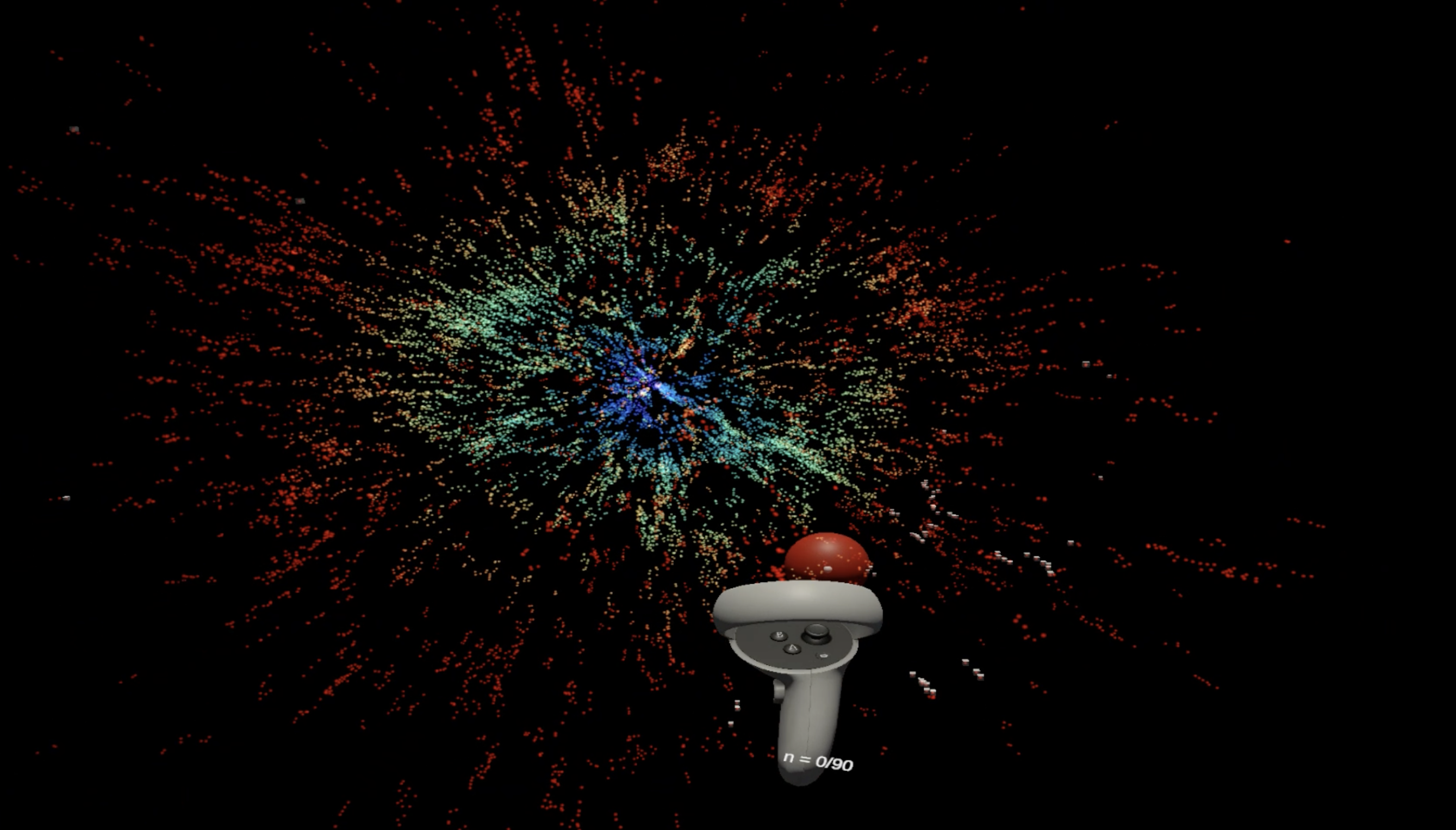}
    \caption{iDaVIE-p with the particle dataset, the controller, and the selection brush. A selection counter can be seen on the controller.}
    \label{fig:idaviep_interface}
\end{figure}

\subsection{Measures}\label{subsec43}
We used several standard and custom metrics to assess performance.

For accuracy we used a modified formula that excludes True Negatives (TN): $Acc' = TP / (TP + FP + FN)$. In large catalogues the huge number of non-target particles would inflate standard accuracy to near 100\% because true negatives swamp everything else \citep{Metz1978, Stehman1997}; our formula measures retrieval of the target set only.

Efficiency was defined as the time to task completion in seconds, capped at 600 seconds (10 minutes). Participants balanced speed and accuracy themselves.

For subjective metrics, we used the SUS \citep{Brooke1996} for a composite 0--100 interface usability score. A score above 68 is considered above average. The NASA-TLX \citep{Hart1988} measures subjective workload across six dimensions (Mental Demand, Physical Demand, Temporal Demand, Performance, Effort, Frustration), with all dimensions weighted equally. The FSS \citep{Jackson1996} measures immersion and flow across nine dimensions: challenge-skill balance, action-awareness merging, clear goals, unambiguous feedback, concentration, sense of control, loss of self-consciousness, transformation of time, and autotelic experience (the activity being experienced as intrinsically rewarding, done for its own sake).

\subsection{Statistical Analysis}\label{subsec44}
Before comparing groups, we checked whether the scores were normally distributed, using the Shapiro-Wilk test, which suits small samples, D'Agostino's K-squared test, and plots of the data. Normality was rejected (Shapiro-Wilk $p<0.05$) for accuracy on both platforms, for desktop completion time, and for the VR distributions of SUS and FSS; only NASA-TLX passed on both platforms. Because at least one of the two platform distributions was non-normal for four of the five measures, and to keep the analysis and effect-size reporting uniform, we used non-parametric Wilcoxon Signed-Rank tests for all paired comparisons, with significance at $\alpha = 0.05$. With only two conditions (VR and Desktop), no overall test is needed before the pairwise one; tests such as Friedman's are for three or more conditions. The subscale analyses in Section~\ref{subsec52} do involve multiple comparisons, so within each family of tests (the five primary measures, the six NASA-TLX subscales, and the nine FSS subscales) we applied a Holm--Bonferroni correction. We report uncorrected $p$-values below and mark which results survive the correction: all four significant primary measures and all five significant NASA-TLX subscales do, whereas among the nine FSS subscales only three do. Effect sizes were calculated using Cohen's $d$ with the pooled standard deviation, and all standard deviations are sample standard deviations. We interpret $d$ by the conventional bands: $|d| \approx 0.2$ small, $|d| \approx 0.5$ medium, $|d| \approx 0.8$ large, and $|d| \geq 1.2$ very large \citep{Cohen1988, Sawilowsky2009}. The sign of $d$ reflects the direction of the paired difference (Desktop minus VR), which is metric-specific: negative values on usability and flow indicate VR scored higher (better), while positive values on workload and completion time indicate Desktop scored higher (worse).

A sensitivity power analysis indicates what this sample can detect. With $N=34$ paired observations, two-tailed $\alpha=0.05$ and power $0.80$, the smallest detectable effect is $d\approx0.50$ for a paired $t$-test and $d\approx0.51$ for the Wilcoxon signed-rank test, which needs about 5\% more participants than a $t$-test to detect the same effect \citep{Cohen1988}. The study was therefore powered to detect medium and larger effects. The effects observed for the significant measures ($|d|$ from $0.90$ to $1.78$) lie well above this threshold, whereas the accuracy effect ($d=0.15$) would have needed roughly $360$ participants to detect at the same power, so the null result on accuracy should be read as ``no medium or large difference'' rather than as evidence of exact equivalence.

\section{Results}\label{sec5}

The results cover the performance of all 34 participants across both platforms. A summary of the descriptive and inferential statistics is given in Table~\ref{tab:results_summary}.

\begin{table}[ht]
\centering
\begin{tabularx}{\textwidth}{l *{2}{X} c c c}
\toprule
\textbf{Metric} & \textbf{Desktop (Mean $\pm$ SD)} & \textbf{VR (Mean $\pm$ SD)} & \textbf{$W$} & \textbf{$p$-value} & \textbf{Effect size ($d$)} \\
\midrule
Accuracy (\%) & $93.09 \pm 8.57$ & $91.61 \pm 11.27$ & 270.5 & 0.858 & 0.148 \\
\addlinespace
Efficiency (s) & $488.29 \pm 113.93$ & $374.52 \pm 129.66$ & 113.0 & \textbf{0.001} & \textbf{0.932} \\
\addlinespace
Usability & $54.04 \pm 20.55$ & $84.63 \pm 12.94$ & 40.5 & \textbf{$<$0.001} & \textbf{$-$1.781} \\
\addlinespace
Workload & $34.68 \pm 18.16$ & $20.42 \pm 13.11$ & 105.5 & \textbf{0.001} & \textbf{0.901} \\
\addlinespace
Flow & $70.21 \pm 12.26$ & $83.15 \pm 8.43$ & 71.5 & \textbf{$<$0.001} & \textbf{$-$1.230} \\
\bottomrule
\end{tabularx}
\caption{Summary of experimental results comparing Desktop and VR (mean $\pm$ sample SD; Wilcoxon signed-rank $W$; Cohen's $d$ with pooled SD, Desktop minus VR). Significant p-values and their corresponding strong effect sizes are in bold.}
\label{tab:results_summary}
\end{table}
\FloatBarrier

\subsection{Accuracy and Efficiency}\label{subsec51}
Accuracy was high on both platforms and a Wilcoxon Signed-Rank test showed no significant difference ($W=270.5$, $p=0.858$, $d=0.148$), so we retain $H_0$(e): VR gives accuracy comparable to desktop. The slightly higher standard deviation in VR ($11.27$ vs.\ $8.57$) could reflect heterogeneity in prior VR experience, so we split VR accuracy by self-reported experience (``novice'', little-to-no prior use, $n=26$; ``experienced'', occasional or frequent use, $n=8$). A Mann-Whitney U test showed no difference between the two groups ($U=99.5$, $p=0.870$), their VR scores were similarly spread out (Levene's test, $W=0.12$, $p=0.727$), and, as a check, experience made no difference on the desktop either ($U=85.5$, $p=0.462$). The higher VR spread cannot be attributed to experience, although the small experienced subgroup limits the power to detect such an effect.

\begin{figure}[tb]
    \centering
    \includegraphics[width=\linewidth]{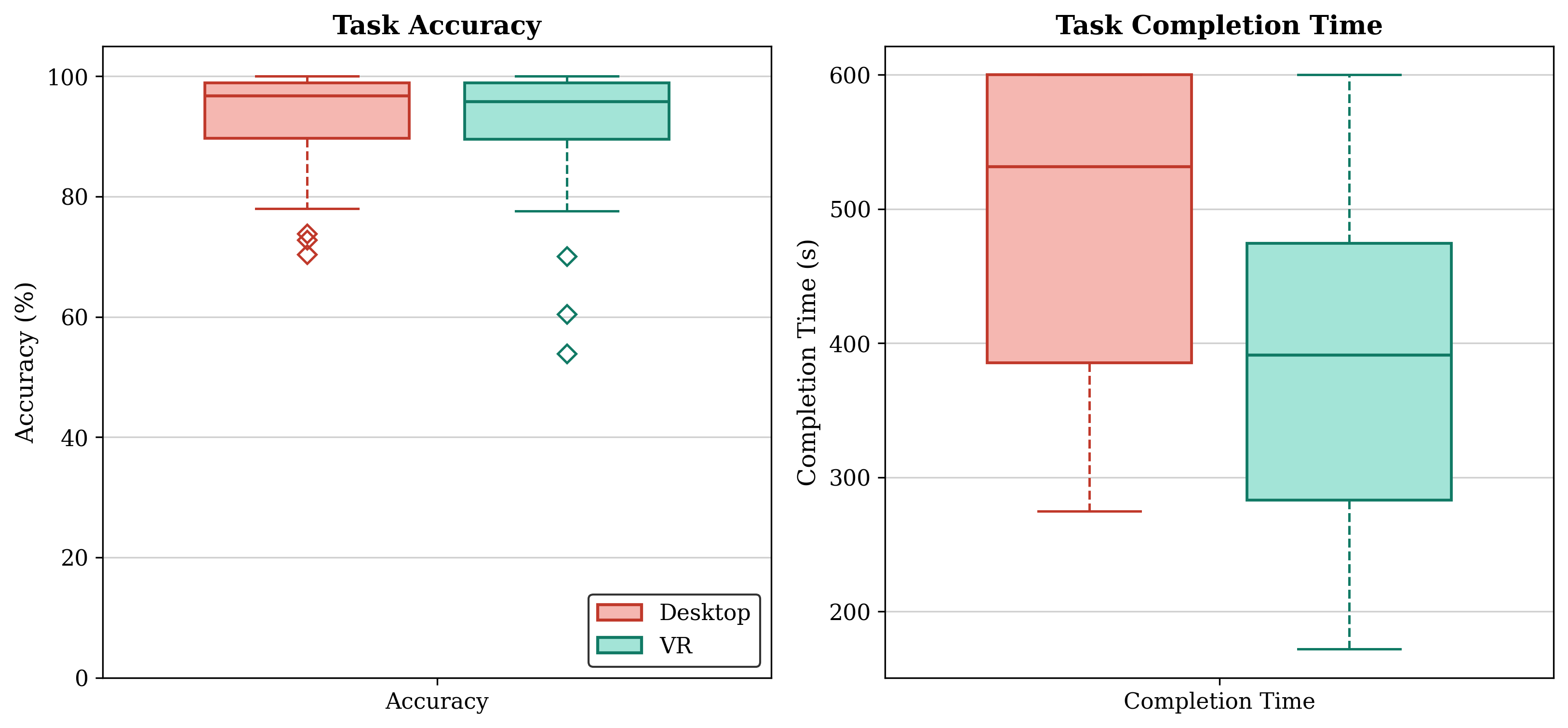}
    \caption{Accuracy (left) and completion time (right) compared between VR and Desktop.}
    \label{fig:accuracy_efficiency}
\end{figure}

The distribution itself (Figure~\ref{fig:accuracy_efficiency}) locates the spread: both platforms cluster at the upper end (medians near 96\%, interquartile ranges of roughly 90--99\%), but the three desktop outliers sit between 70\% and 74\% whereas the three VR outliers lie near 70\%, 60\%, and 54\%. A handful of VR participants therefore struggled more acutely than any desktop participant, raising the VR standard deviation without shifting the median; plausible causes are disorientation with VR locomotion or aggressive deselection that could not be corrected within the time cap. The null result on experience also has a positive side: VR performance was robust to prior familiarity with the medium, which matters where most users have no VR training.

Efficiency, by contrast, showed a substantial and significant advantage for VR: users were about 23.3\% faster on average ($W=113.0$, $p=0.001$, $d=0.932$). The completion-time distribution in Figure~\ref{fig:accuracy_efficiency} reveals a further asymmetry: the slowest quarter of desktop participants all hit the 600-second limit: 11 of 34 desktop participants ran out of time, against one VR participant. Their real times would have been longer, so 23\% most likely understates the gap. The same experience-based split on VR completion time was non-significant ($U=102.0$, $p=0.952$; novice and experienced means 374.2~s and 375.5~s), so the gain is not an artefact of a few proficient VR users.

\subsection{Usability, Workload, and Flow}\label{subsec52}
Subjective metrics overwhelmingly favoured the immersive VR environment. On the SUS, VR scored significantly higher ($W=40.5$, $p<0.001$, $d=-1.781$). The VR mean of 84.63 places it in the ``Excellent'' usability category, while the desktop mean of 54.04 is below average (``OK'').

Participants also reported significantly lower workload in VR on the NASA-TLX ($W=105.5$, $p=0.001$, $d=0.901$), a 41\% reduction. The subscales (Table~\ref{tab:tlx_subscales}) show where this comes from: five of six dimensions moved in VR's favour and all five survive the Holm correction, led by Frustration ($-77$\%, $d=1.08$) and Temporal Demand ($-59$\%, $d=0.78$), followed by Effort ($-42$\%, $d=0.75$), Performance ($-53$\%, $d=0.54$), and Mental Demand ($-27$\%, $d=0.50$). Physical Demand alone moved against VR ($+32$\%, $p=0.098$, $d=-0.28$, not significant), consistent with a brush metaphor that involves reaching and sweeping. The workload benefit of VR is thus primarily about how the task feels to execute (less frustrating, less rushed, with a stronger sense of having performed well) rather than a simple reduction in cognitive effort, and the modest physical cost is more than offset.

Flow was also significantly higher in VR ($W=71.5$, $p<0.001$, $d=-1.230$). All nine FSS dimensions moved in VR's favour (Table~\ref{tab:fss_subscales}), but the effect is concentrated in three of them, which are the only ones that survive the Holm correction. By far the largest is Autotelic Experience ($+69$\%, $d=-2.09$), the degree to which the activity is intrinsically rewarding: every participant rated it equal or higher in VR ($W=0$). Sense of Control ($+26$\%, $d=-1.09$) and Action--Awareness Merging ($+28$\%, $d=-0.74$) follow. Sense of Control is the subscale most directly tied to the design argument in Section~\ref{subsec32}: in VR the selection sits exactly where the hand is, so each movement has a predictable result, whereas the desktop box makes the user map how 2D mouse input acts in 3D, and its large advantage suggests that participants felt this difference. The remaining six subscales show small-to-medium differences ($|d|$ from 0.31 to 0.62) that do not survive correction; Clear Goals in particular is high on both platforms ($88$ vs.\ $92$), as expected for a task with explicitly marked targets. Autotelic Experience points beyond task performance: participants found VR enjoyable in a way desktop was not, which matters for sustained engagement over long sessions.

\begin{figure}[tb]
    \centering
    \includegraphics[width=\linewidth]{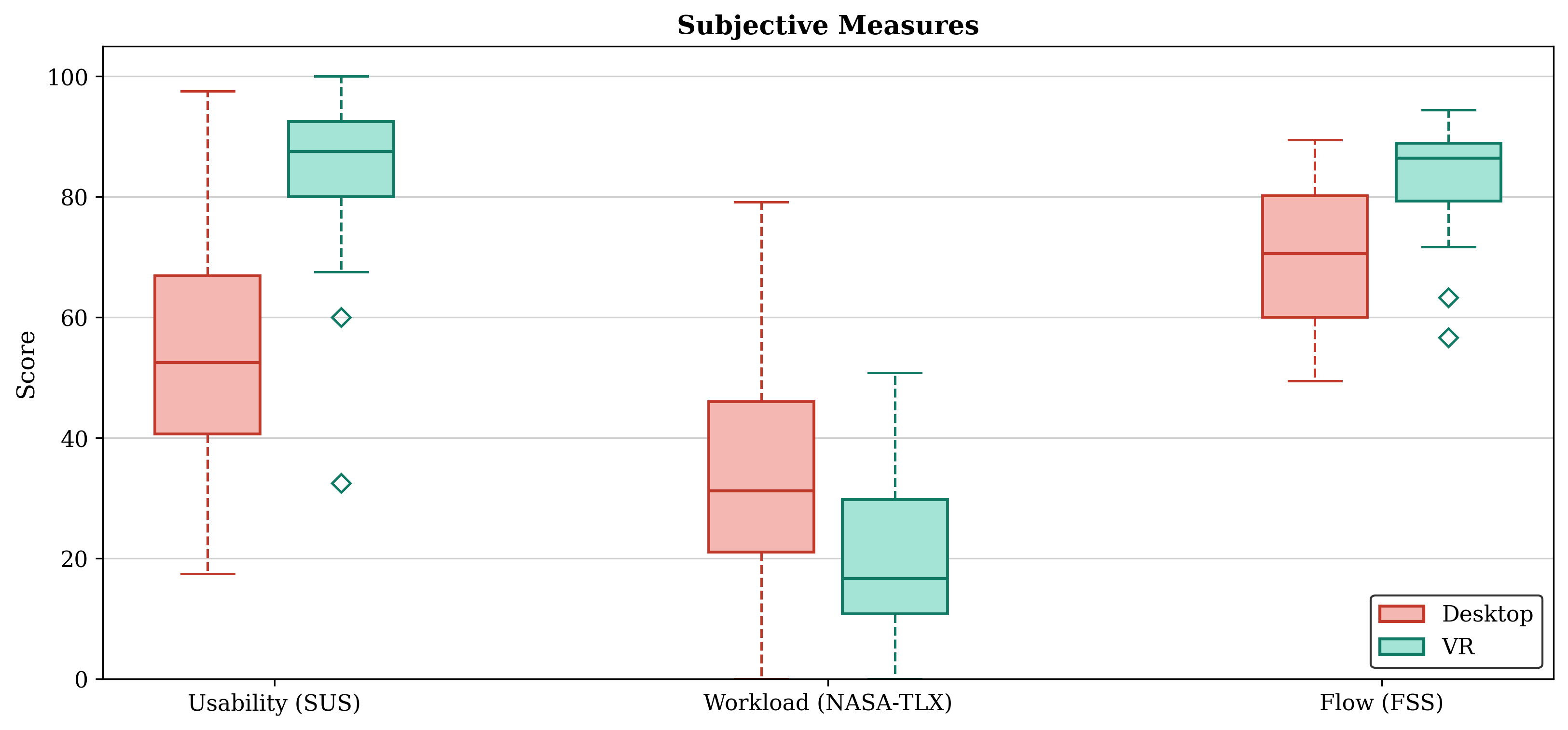}
    \caption{Subjective measures compared between VR and Desktop.}
    \label{fig:subjective_metrics}
\end{figure}

A consistent minority pattern appears across all three subjective measures (Figure~\ref{fig:subjective_metrics}): VR distributions are tighter and better than desktop, but the usability and flow panels each contain two VR outliers well below the main cluster, and the workload panel has none. In Figures~\ref{fig:accuracy_efficiency} and~\ref{fig:subjective_metrics}, a data point counts as an outlier if it lies more than 1.5 times the interquartile range (the height of the box) above or below the box. The lowest VR SUS score ($\sim$32) is below the desktop median, and FSS shows two low outliers near 57 and 63, whereas the desktop distributions are broad but have no flagged outliers. Together with the accuracy outliers in Section~\ref{subsec51}, this suggests VR produces a more polarised response: most participants found it substantially better, but two of the 34 participants found it meaningfully worse on both objective and subjective measures: they produced two of the three lowest VR accuracy scores and the two lowest VR usability ratings (Section~\ref{sec7}).

\begin{table}[!ht]
\centering
\begin{tabularx}{\textwidth}{l *{2}{X} c c c c}
\toprule
\textbf{Subscale} & \textbf{Desktop (Mean $\pm$ SD)} & \textbf{VR (Mean $\pm$ SD)} & \textbf{$W$} & \textbf{$p$-value} & \textbf{$d$} & \textbf{\% change} \\
\midrule
Mental Demand    & $48.7 \pm 25.0$ & $35.7 \pm 26.8$ & 116.5 & \textbf{0.017} & \textbf{$+0.50$} & $-27$\% \\
\addlinespace
Physical Demand  & $20.0 \pm 22.7$ & $26.5 \pm 24.0$ & 110.5 & 0.098 & $-0.28$ & $+32$\% \\
\addlinespace
Temporal Demand  & $32.2 \pm 30.1$ & $13.1 \pm 17.2$ & 28.0 & \textbf{$<$0.001} & \textbf{$+0.78$} & $-59$\% \\
\addlinespace
Performance      & $28.1 \pm 31.5$ & $13.1 \pm 23.4$ & 81.0 & \textbf{0.005} & \textbf{$+0.54$} & $-53$\% \\
\addlinespace
Effort           & $45.1 \pm 27.5$ & $26.3 \pm 22.5$ & 101.5 & \textbf{0.004} & \textbf{$+0.75$} & $-42$\% \\
\addlinespace
Frustration      & $34.0 \pm 31.2$ & $7.8 \pm 14.0$ & 46.0 & \textbf{$<$0.001} & \textbf{$+1.08$} & $-77$\% \\
\bottomrule
\end{tabularx}
\caption{NASA-TLX subscale breakdown, VR vs.\ Desktop. Positive $d$ indicates Desktop scored higher (VR reduced workload); negative $d$ indicates VR scored higher (applies only to Physical Demand). Paired Wilcoxon Signed-Rank tests, $N=34$; $d$ uses the pooled SD. Uncorrected p-values; all five significant subscales remain significant after Holm--Bonferroni correction. Significant p-values and medium-or-larger effect sizes are in bold.}
\label{tab:tlx_subscales}
\end{table}

\begin{table}[!ht]
\centering
\begin{tabularx}{\textwidth}{l *{2}{X} c c c c}
\toprule
\textbf{Subscale} & \textbf{Desktop (Mean $\pm$ SD)} & \textbf{VR (Mean $\pm$ SD)} & \textbf{$W$} & \textbf{$p$-value} & \textbf{$d$} & \textbf{\% change} \\
\midrule
Challenge-Skill Balance    & $75.4 \pm 16.5$ & $82.9 \pm 13.7$ & 142.0 & 0.037 & $-0.50$ & $+10$\% \\
\addlinespace
Action-Awareness Merging   & $52.2 \pm 20.2$ & $66.8 \pm 19.4$ & 106.5 & \textbf{0.003} & \textbf{$-0.74$} & $+28$\% \\
\addlinespace
Clear Goals                & $87.9 \pm 13.9$ & $91.8 \pm 10.6$ & 121.0 & 0.261 & $-0.31$ & $+4$\% \\
\addlinespace
Unambiguous Feedback       & $77.5 \pm 17.4$ & $84.7 \pm 14.2$ & 137.5 & 0.050 & $-0.45$ & $+9$\% \\
\addlinespace
Concentration              & $84.7 \pm 15.9$ & $92.6 \pm 8.9$ & 62.5 & 0.021 & $-0.62$ & $+9$\% \\
\addlinespace
Sense of Control           & $70.7 \pm 17.6$ & $89.0 \pm 15.8$ & 59.0 & \textbf{$<$0.001} & \textbf{$-1.09$} & $+26$\% \\
\addlinespace
Loss of Self-Consciousness & $77.8 \pm 21.9$ & $87.9 \pm 16.7$ & 86.0 & 0.023 & $-0.52$ & $+13$\% \\
\addlinespace
Transformation of Time     & $50.0 \pm 16.9$ & $58.7 \pm 22.1$ & 129.5 & 0.034 & $-0.44$ & $+17$\% \\
\addlinespace
Autotelic Experience       & $55.6 \pm 23.6$ & $94.0 \pm 10.8$ & 0.0 & \textbf{$<$0.001} & \textbf{$-2.09$} & $+69$\% \\
\bottomrule
\end{tabularx}
\caption{Flow State Scale subscale breakdown, VR vs.\ Desktop. Subscales are scored from the four FSS items of each dimension (mean item score scaled to 0--100). Negative $d$ indicates VR scored higher; $d$ uses the pooled SD. Paired Wilcoxon Signed-Rank tests, $N=34$, uncorrected p-values; bold marks the three subscales that remain significant after Holm--Bonferroni correction.}
\label{tab:fss_subscales}
\end{table}

\section{Discussion}\label{sec6}

The experiment largely favoured VR, and we reject $H_0$ for usability (a), workload (b), flow (c), and completion time (d). Accuracy (e) was statistically comparable between platforms, so we retain $H_0$ for that measure. On balance, both user experience and performance efficiency were superior in VR.

Despite the novelty of VR for most participants, tasks were completed 23\% faster and with 41\% lower perceived workload. Restricting the analysis to the 26 participants with little or no prior VR experience leaves every significant difference intact (completion time $p=0.004$, SUS $p<0.001$, NASA-TLX $p=0.003$, FSS $p<0.001$), while accuracy stays non-significant ($p=0.638$). This comes down to the nature of the task and the data. A particle catalogue is 3D, so selecting a subset of it is a 3D spatial task; on a desktop that task is abstracted through 2D input on a 2D projection, which demands constant effort to interpret depth cues and manage camera angles. In VR the brush simply follows the hand: to select a cluster of galaxies, the user reaches out and moves a hand through it, with no mental translation between 2D input and 3D data.

The large gap in usability (84.63 vs.\ 54.04) highlights the limits of desktop metaphors for volumetric selection: precise for single targets, cumbersome for bulk 3D operations. The multi-step desktop box (position, scale, verify depth, confirm) contrasts with the immediate VR brush, and most desktop participants in fact gravitated to single-particle clicking, reporting that it felt faster than positioning and confirming a box for each cluster. Rather than use the tool designed for bulk selection, they broke the task into many single clicks, which shows how high the overhead of specifying a 3D volume through a 2D interface is. The VR brush makes bulk selection the path of least resistance, which is part of why the efficiency gap is as large as it is.

The speed/accuracy trade-off left to participants in Section~\ref{subsec43} did not materialise: VR participants were 23\% faster while accuracy was statistically indistinguishable (91.6\% vs.\ 93.1\%, $p=0.858$). A 1.5-point difference on a measure whose interquartile range spans about nine points is far smaller than the loss one would expect had users traded precision for speed, so the more plausible reading is that the direct 3D mapping removed overhead rather than encouraging haste. This matters because in real workflows a false positive in the analysis set can be costly, so a method that gained speed by losing accuracy would be unattractive.

More broadly, many VR applications adapt essentially 2D tasks to 3D (typing on a virtual keyboard, for example). Particle selection is the opposite: an intrinsically 3D task that the desktop must squeeze into 2D. This inversion helps explain why even participants inexperienced in VR performed better in the immersive environment.

Neither the task nor the result is astronomy-specific. Isolating a spatially coherent subset of points, refining it, and coping with occlusion recurs wherever data is an unstructured point cloud, from anatomical point clouds derived from MRI \citep{Beetz2023} to LiDAR surveys segmented into structural components \citep{Vacca2023}. Each shares the feature that made VR advantageous here: the data is intrinsically 3D, and a 2D interface forces the user to express a 3D selection through operations on a flat projection. We therefore expect the qualitative pattern to carry over, with a magnitude that depends on the data, as the limitations below discuss.

\section{Limitations}\label{sec7}
Several limitations should be kept in mind when interpreting these results.

The study used a single dataset. The 2MRS consists of roughly spherical galaxy clusters spread through open 3D space, close to the shape a spherical brush is designed for and shared by much of the data we expect the tool to be used on: galaxy and source catalogues, the particle output of cosmological simulations such as IllustrisTNG \citep{Nelson2019TNG} and SIMBA \citep{Dave2019SIMBA}, and compact groups of point features in other domains such as microscopy. Other layouts will change the size of the advantage: long thread-like structures (the cosmic web at larger scales, branching blood vessels) suit stroke- or line-based tools better than a round brush, and datasets with strong density contrast (a dense LiDAR wall next to sparse terrain) favour a brush adapted to local density rather than one whose radius is set by the user. The direction of the result is unlikely to reverse, because the desktop's core disadvantage of projecting 3D into 2D remains, but the magnitude may shrink, and measuring this across data types is a priority for future work.

The study design has three further constraints. All participants came from a single institution and mostly from two departments, which matches the target audience but limits generalisation. Platform order was counterbalanced but target-set order was not: Set A was always the first task (Section~\ref{subsubsec423}), so differences between the sets are confounded with task order, although the 15-minute VR playground will have offset part of any learning effect. Simulator sickness was not measured with an instrument such as the Simulator Sickness Questionnaire; no participant reported discomfort, but mild effects cannot be ruled out.

The desktop box is a deliberately plain baseline. Smarter desktop techniques guess the intended region from a lasso or click, using how densely the particles are packed in CAST \citep{Yu2016} or a neural network trained on viewpoints and lassos in LassoNet \citep{Chen2019LassoNet}, and are substantially faster than plain cylinder selection at similar or better accuracy, so some of the 23\% advantage could plausibly be closed on the desktop side. They solve a different problem, however: they make a 3D selection easier to describe with a 2D gesture, whereas the VR brush removes the 2D gesture altogether, and the same density adaptation could equally be added to the VR brush. A three-way comparison between the VR brush, a plain desktop box, and such a smarter desktop selector, with and without density adaptation in VR, would separate the benefit of smarter interaction from the benefit of immersion.

\section{Conclusions}\label{sec8}
This study shows that VR significantly outperforms a traditional desktop interface for 3D particle selection in astronomy. Accuracy is maintained, while VR offers higher efficiency (23\% faster), higher usability (84.63 vs.\ 54.04 SUS), lower workload (41\% reduction), and deeper flow (83.15 vs.\ 70.21 FSS), and these benefits held even for participants with little VR experience. We attribute this to the direct match between the user's hand movements and the 3D data, which removes the mental translation from 2D input to 3D data that desktop interfaces require and makes the interaction faster, more satisfying, and less taxing. VR is therefore a viable and advantageous tool for analysing particle-oriented astronomical data, and we advocate its wider adoption in scientific settings where tasks and datasets are inherently three-dimensional.

\backmatter

\bmhead{Acknowledgements}
This work made use of the Inter-University Institute for Data Intensive Astronomy (IDIA) visualisation lab (UCT/IDIA vislab). IDIA is a partnership of the University of Cape Town, the University of Pretoria and the University of the Western Cape. This work made use of the iDaVIE (Immersive Data Visualisation Interactive Explorer) software in its particle rendering mode.

\section*{Declarations}

\bmhead{Funding}
This project has received funding from the European Research Council (ERC) under the European Union's Seventh Framework Programme (FP/2007-2013)/ERC (grant agreement no. 291531, “HIStoryNU”) and the Horizon 2020 research and innovation programme (grant agreement no. 679627, ``FORNAX''), from MAECI (Italian Ministry of Foreign Affairs and International Cooperation) Grant Numbers PGR ZA23GR03 (RADIOMAP) and PGR ZA18GR02 (RADIOSKY2020), from DSTNRF Grant Number 113121 as part of the ISARP Joint Research Scheme, from ISARP RADIOMAP Joint Research Scheme (DSI-NRF Grant Number 150551), from the CPRR Projects (DSI-NRF Grant Number SRUG2204254729) and the Italian National Institute for Astrophysics and from the Inter-University Institute for Data Intensive Astronomy (IDIA).

\bmhead{Competing interests}
The authors declare no competing interests.

\bmhead{Ethics approval}
Ethics clearance was obtained from the University of Cape Town.

\bmhead{Consent to participate}
All participants gave written informed consent before the session, including consent to the use of anonymised data for research and publication.

\bmhead{Data availability}
The anonymised per-participant data (task scores, all questionnaire item responses and the two target sets), a codebook, and a script that reproduces every statistic in this article are available at \url{https://huggingface.co/datasets/boazkg/iDaVIE-particle-selection-vr-vs-desktop}.

\bibliography{sn-bibliography}

\end{document}